\documentclass[10pt,conference]{IEEEtran}
\IEEEoverridecommandlockouts

\usepackage{cite}
\usepackage{amsmath,amssymb,amsfonts}
\usepackage{graphicx}
\usepackage{textcomp}
\usepackage{xcolor}

\usepackage[ruled,vlined,linesnumbered]{algorithm2e}
\usepackage{tcolorbox}
\usepackage{float}
\usepackage{algpseudocode}

\def\BibTeX{{\rm B\kern-.05em{\sc i\kern-.025em b}\kern-.08em
    T\kern-.1667em\lower.7ex\hbox{E}\kern-.125emX}}
\begin{document}

\title{From Ripple to Boundary: Deterministic Impact Boundaries for LLM-Based Software Evolution

}


\author{
\IEEEauthorblockN{
    Jiameng Bai\textsuperscript{1},
    Zhitao Wang\textsuperscript{2,*},
    Long Wang\textsuperscript{2},
    Keyue Qian\textsuperscript{1},
    Sai Wu\textsuperscript{1,*}\\
}
\IEEEauthorblockA{
    \textsuperscript{1}Zhejiang University, Hangzhou, China\\
    \textsuperscript{2}WeChat Pay, Tencent, Shenzhen, China\\
    \{baijiameng,kyqian,wusai\}@zju.edu.cn\\
    \{zhitaowang,oliverlwang\}@tencent.com
}

\thanks{\textsuperscript{*}Corresponding authors.}
}


\maketitle

\begin{abstract}
The landscape of \textbf{software evolution} is undergoing a fundamental paradigm shift in the era of Large Language Models (LLMs). As industrial software engineering increasingly transitions toward design-driven code generation, the bottleneck of evolution has shifted from writing code to maintaining \textbf{deterministic consistency} between multi-source design assets (e.g., sequence diagrams, class diagrams) and the generated code. When a design asset changes, LLM agents currently lack a deterministic impact boundary to identify which code must be evolved, leading to two failure modes: hallucinating missing changes or breaking unrelated code. Probabilistic approaches—such as semantic retrieval or text-based diffing—fail to capture the structural ripple effects across highly interconnected design dependencies. 

To address this, we propose a paradigm shift from probabilistic semantic retrieval to \textbf{deterministic structural constraints}. We formalize the design assets as a Knowledge Graph (KG) and organize them into a content-addressed Merkle Directed Acyclic Graph (DAG). Our key insight reveals a \textbf{topological isomorphism} between the logical ripple effect of a design change and the content hash propagation within the DAG. By leveraging this, we transform ambiguous impact analysis into \textbf{verifiable structural operations}. Through a two-phase diffing process, we extract a precise \textit{causal chain} that defines the exact modification boundary. This boundary enables \textbf{deterministic context injection}, providing the LLM agent with strictly scoped code and design updates, thereby fundamentally guaranteeing evolution safety. As a by-product, the KG serves as an agent-queryable knowledge base for natural language design retrieval. Early deployment on a large-scale industrial payment platform demonstrates that our mechanism successfully computes the exact impact boundary and generates incremental code modifications that meet the expected design changes.

\end{abstract}


\begin{figure*}[htbp!]
\centering
\includegraphics[width=0.9\textwidth]{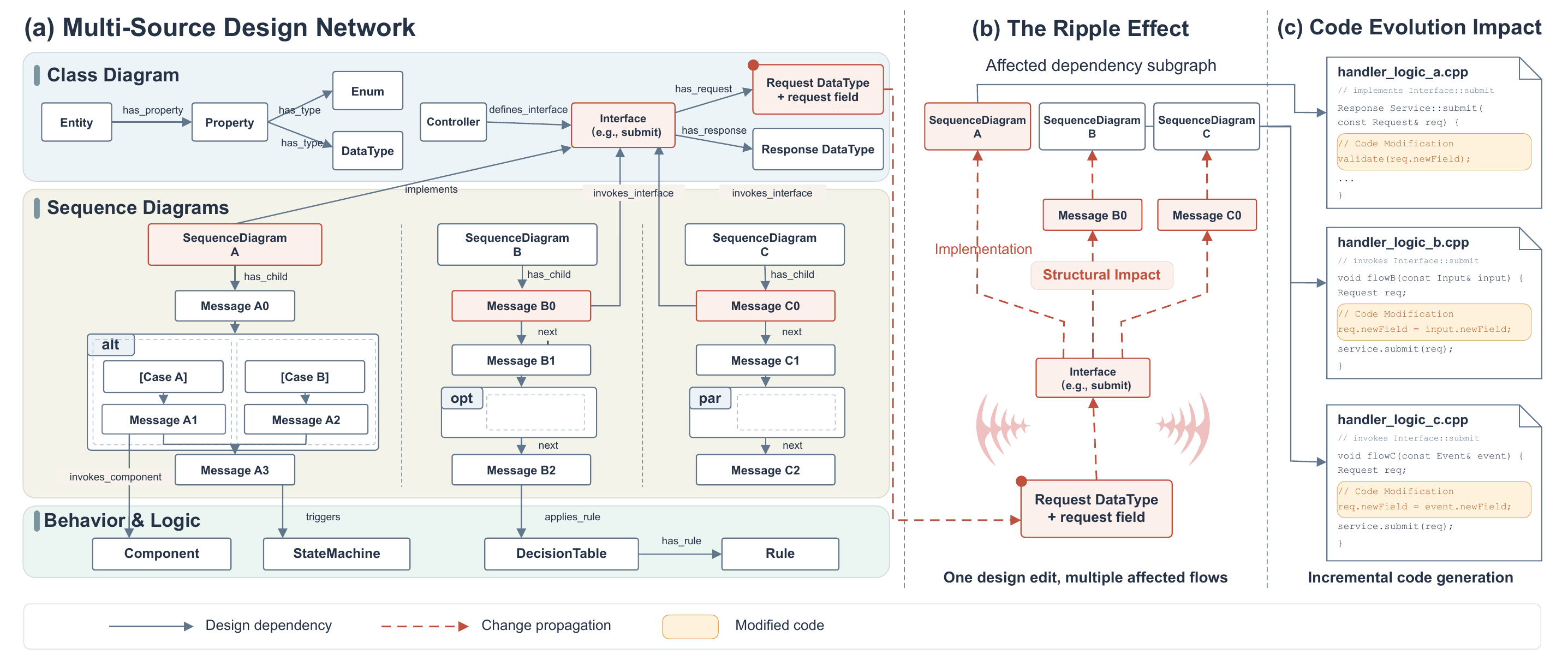}
\caption{The Design Ripple Challenge: (a) A multi-source design network across class diagrams, sequence diagrams, and behavior/logic layers; (b) A single field edit in the Request DataType triggers a structural ripple effect across multiple sequence diagrams; (c) The ripple necessitates incremental code generation.}
\label{fig:ripple}
\vspace{-1em}
\end{figure*}

\section{Introduction}

The landscape of \textbf{software evolution} is undergoing a fundamental paradigm shift. In our domain of industrial payment software development, we have transitioned toward a \textbf{design-driven code generation} paradigm. To make this concrete, consider an API-based transaction system: a sequence diagram captures a specific business use case, where each \textit{Message} corresponds to an API or function call, often augmented with decision rule tables. Instead of manual coding, developers define these multi-source design assets (sequence diagrams, class diagrams, and rule tables) as the single source of truth, from which LLM agents synthesize and evolve the actual code.

However, as the design asset base scales, this evolutionary paradigm exposes a critical vulnerability during incremental iterations: when a design changes, the LLM agent lacks a \textbf{deterministic impact boundary}. Consider a scenario: a core interface's signature (e.g., its function name or request structure) is modified. Since this interface is invoked by dozens of messages across multiple sequence diagrams, this single logical change triggers a massive \textit{ripple effect}. In a design-to-code setting, every affected sequence diagram dictates that its corresponding generated code must be evolved accordingly. Traditional approaches fail to capture this structural propagation. Text-based version control operates on physical text, losing logical dependencies. Even when leveraging LLMs to deduce the impact based on semantic similarity or RAG, the results remain probabilistic. Fundamentally, \textbf{LLMs inherently lack the capability to perform deterministic transitive closure calculations over complex graph structures}, making them prone to missing deep cascaded dependencies. This boundary ambiguity leads to two failure modes: \textit{under-modification} (hallucinating or missing necessary code changes) and \textit{over-modification} (breaking unrelated shared code).

To address this, we propose a paradigm shift: \textbf{moving from probabilistic semantic retrieval to deterministic structural constraints}. Because design assets are highly interconnected (e.g., a message invokes an interface defined in a class diagram), we model them as a directed graph and store them in a \textbf{Knowledge Graph (KG)} to explicitly capture their complex logical dependencies. Building upon this KG, our key insight is to organize it into a content-addressed \textbf{Merkle Directed Acyclic Graph (DAG)}. Inspired by the underlying architecture of Git, each design node is stored via shallow serialization, and its content hash is derived from its own properties and its children's hashes. This structural representation makes the logical ripple effect computable: any leaf-level change inherently alters the hashes of its ancestors. By comparing the hashes of new and old snapshots, our two-phase diffing process efficiently prunes unchanged subtrees and drills down into the modified ones, yielding a precise \textbf{causal chain of impact} (e.g., ``Interface X is affected because its request structure changed, consequently affecting Sequence Y'').

Crucially, instead of relying on probabilistic agentic search during generation, our system leverages this computed causal chain to perform \textbf{deterministic context injection}. The agent is provided with a pre-packaged specification containing the precise causal chain, the newly fetched design assets, and the existing code, all strictly confined within the computed boundary. This rigid structural constraint eliminates the need for the LLM to ``search'' for context, ensuring that incremental code evolution is restricted to truly affected regions. 

Beyond automated evolution, the constructed KG yields a valuable by-product: it serves as an \textbf{agent-queryable knowledge base}. During the design phase, when developers need to understand complex structural dependencies (e.g., ``Which sequences invoke this interface?''), an LLM agent can safely traverse the graph via natural language and multi-step Cypher queries. This unlocks the design assets from static documents into an interactive, dynamic engineering asset, assisting human designers in their daily workflows.

We have deployed this mechanism on a large-scale industrial payment microservice platform. Early evidence from real-world flow modifications demonstrates that our approach effectively constrains the agent's generative process, ensuring that code changes are precisely scoped and safe.
By treating design changes as hash propagation over a Merkle DAG, we shift LLM-based software evolution from probabilistic context retrieval to deterministic structural enforcement.


\begin{figure*}[t]
\centering
\includegraphics[width=0.9\textwidth]{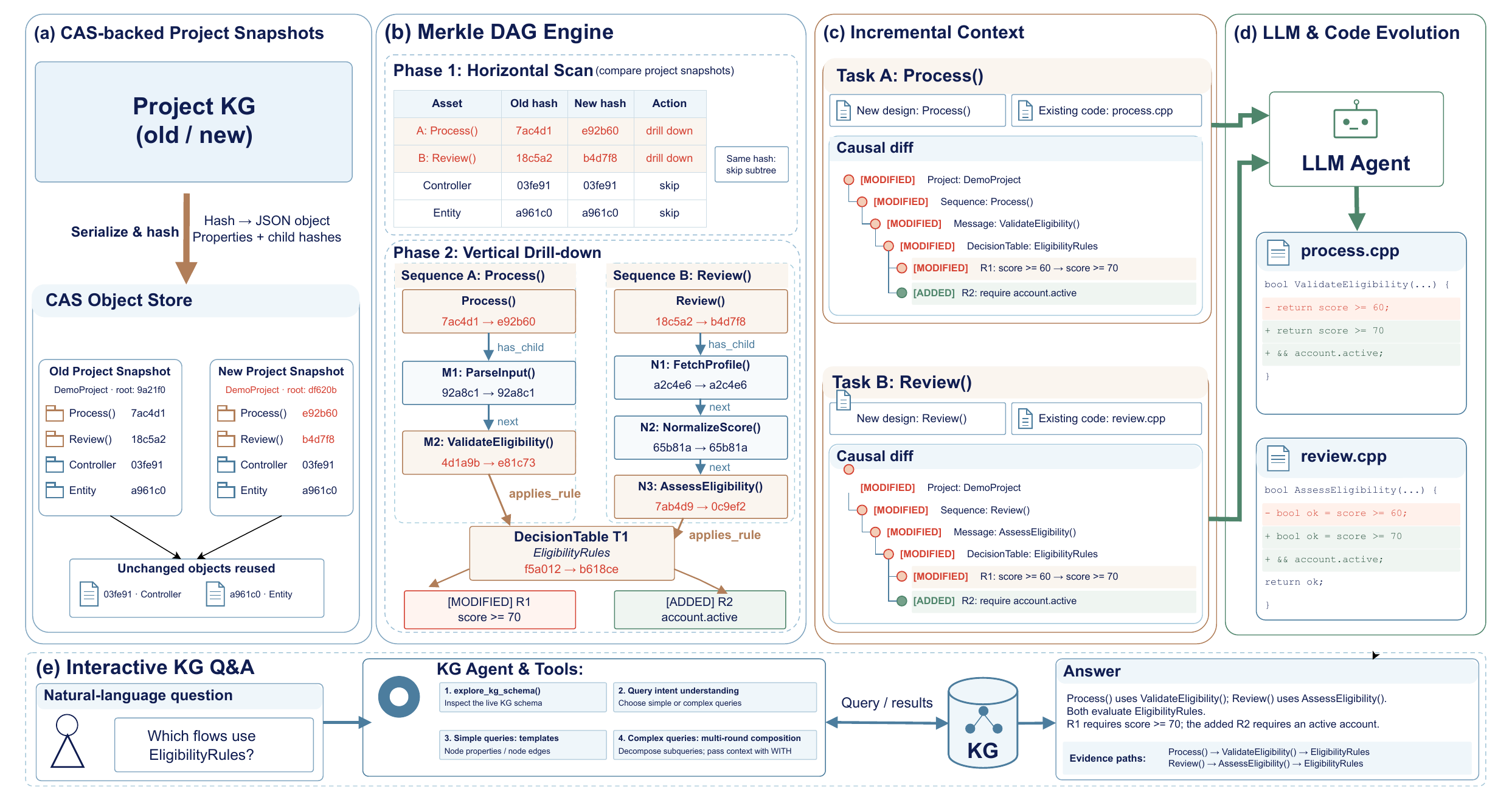}
\caption{The proposed workflow. (a) CAS-backed snapshots; (b) Two-phase Diff over Merkle DAG; (c) Per-sequence incremental context; (d) Constrained code evolution; (e) Interactive KG Q\&A.}
\label{fig:approach}
\vspace{-1em}
\end{figure*}

\section{The Design Ripple in Software Evolution}

\textbf{A. The Multi-Source Design Network.}
In our industrial workflow, engineers primarily focus on drawing sequence diagrams to define business flows, while concurrently editing other heterogeneous assets such as class diagrams and rule tables. 
To formally manage these assets, we integrate them based on internal business rules, define their structural relationships, and store them in a \textbf{KG}. Figure \ref{fig:ripple}(a) illustrates a \textit{simplified} schema of this KG (the actual network is far denser). 

The KG spans three primary layers. First, the \textbf{Class Diagram layer} defines static structures, where an Interface encapsulates complex Request and Response DataTypes. Second, the \textbf{Sequence Diagram layer} orchestrates flows through Messages that invoke these interfaces. Third, the \textbf{Behavior, Logic \& Infrastructure layer} manages dynamic business rules (StateMachines, DecisionTables) and encapsulates shared infrastructure capabilities (e.g., login component) invoked by Messages. Because these layers are heavily cross-referenced (e.g., a single Interface is invoked by multiple Messages across different Sequence Diagrams), they form a complex Directed Acyclic Graph (DAG) rather than a simple tree.

\textbf{B. The Ripple Effect and Boundary Ambiguity.}
The critical challenge for software evolution arises when \textbf{any arbitrary asset} within this dense graph is modified, rather than just a specific leaf node. Consider the scenario in Figure \ref{fig:ripple}(a): a new field is added to the \textbf{Request DataType} of the \texttt{submit} interface. This single structural edit triggers a massive \textit{ripple effect}. The interface's structural hash is invalidated, cascading to any \texttt{Message} that binds to it (e.g., \texttt{Message B0} and \texttt{C0}). This impact propagates upward, altering the entire \texttt{SequenceDiagram} (e.g., Sequence Diagram B and C). 
As depicted in Figure \ref{fig:ripple}(b), \textbf{one design edit leads to multiple affected flows}. When an LLM agent is tasked with this evolution, it faces the \textbf{Boundary Ambiguity} problem: it must mechanically determine the exact structural boundary of the impacted flows.

\textbf{C. Code Evolution Impact.}
In a design-to-code setting, this design ripple directly mandates code evolution. As shown in Figure \ref{fig:ripple}(c), each affected Sequence Diagram corresponds to specific generated code files (e.g., \texttt{handler\_logic\_b.cpp}). The LLM agent must modify the core implementation (\texttt{handler\_logic\_a.cpp}) and ensure upstream callers handle the new \texttt{Request} field correctly. 
If the structural boundary is not precisely computed, two failure modes occur: \textit{under-modification} (missing callers, leading to compile or runtime errors) and \textit{over-modification} (incorrectly altering unrelated flows). To ensure reliable software evolution, we must answer a fundamental question: \textit{Given a specific design edit, how can we mechanically compute the exact logical boundary of the impacted flows and code?} This requires translating the ambiguous semantic impact into a deterministic structural boundary.

\begin{algorithm}[t]
\caption{Two-Phase Diff for Causal Boundary}
\label{alg:diff}
\KwIn{Old snapshot \(S_{old}\), New snapshot \(S_{new}\)}
\KwOut{Causal Diff Tree \(T_{diff}\)}

\tcp{Phase 1: Horizontal Scan}
\(M_{old}, M_{new} \gets \text{Flatten}(S_{old}), \text{Flatten}(S_{new})\)\;
\(V_{add} \gets M_{new} \setminus M_{old}\)\;
\(V_{del} \gets M_{old} \setminus M_{new}\)\;
\(V_{mod} \gets \{ id \in M_{old} \cap M_{new} \mid H_{old}(id) \neq H_{new}(id) \}\)\;
\BlankLine

\tcp{Phase 2: Vertical Drill-down}
\ForEach{\(id \in V_{mod}\)}{
    \textbf{DrillDown}(\(id, H_{old}(id), H_{new}(id)\))\;
}

\SetKwFunction{FDrillDown}{DrillDown}
\SetKwProg{Fn}{Function}{:}{}
\Fn{\FDrillDown{\(id, H_{old}, H_{new}\)}}{
    \(props_{old} \gets \text{LoadProps}(H_{old})\)\;
    \(props_{new} \gets \text{LoadProps}(H_{new})\)\;
    \If{\(props_{old} \neq props_{new}\)}{
        Record property change as \textbf{root cause} in \(T_{diff}\)\;
    }
    \(children_{old} \gets \text{GetChildren}(H_{old})\)\;
    \(children_{new} \gets \text{GetChildren}(H_{new})\)\;
    
    \ForEach{\(c \in children_{old} \cup children_{new}\)}{
        \uIf{\(c \in children_{new} \setminus children_{old}\)}{
            Record \(c\) as \textbf{ADDED} in \(T_{diff}\)\;
        }
        \uElseIf{\(c \in children_{old} \setminus children_{new}\)}{
            Record \(c\) as \textbf{DELETED} in \(T_{diff}\)\;
        }
        \ElseIf{\(H_{old}(c) \neq H_{new}(c)\)}{
            \textbf{DrillDown}(\(c.id, H_{old}(c), H_{new}(c)\))\;
        }
    }
}
\end{algorithm}

\section{Approach}

The workflow of our approach is illustrated in Figure \ref{fig:approach}. 

\textbf{3.1 Content-Addressable Representation.} To enable deterministic change detection, we model the design Knowledge Graph (KG) as a Merkle DAG. Each node \(n\) is assigned a structural hash recursively defined as:
\begin{equation}
\footnotesize
H(n) = \text{Hash}\Big( \text{Type}(n) \parallel \text{Props}(n) \parallel H(c_1) \parallel \dots \parallel H(c_k) \Big)
\end{equation}
where \(\text{Type}(n)\) is the node label, \(\text{Props}(n)\) are its scalar properties, and \(H(c_i)\) are the hashes of its \(k\) child nodes. We adopt \textbf{shallow serialization}, storing only the \textit{hash pointers} of children. This design guarantees that any change propagates upward to alter the root hash. As shown in Figure \ref{fig:approach}(a), since the file name in CAS is its hash, unchanged subtrees (e.g., the \texttt{Controller}) are naturally reused and stored only once, completely eliminating redundant storage and computation.

\textbf{3.2 Two-Phase Diff for Causal Boundary.} Given the old snapshot \(S_{old}\) and new snapshot \(S_{new}\), the engine extracts the exact causal boundary via a two-phase algorithm (Algorithm \ref{alg:diff}).
Figure \ref{fig:approach}(b) illustrates this process.

In \textbf{Phase 1: Horizontal Scan (Lines 1-4)}, the engine flattens the snapshots into hash maps (Lines 1) and performs set operations to classify assets into added (\(V_{add}\), Line 2), deleted (\(V_{del}\), Line 3), and modified (\(V_{mod}\), Line 4) sets. As shown in Figure \ref{fig:approach}(b), because the hashes of \texttt{Controller} and \texttt{Entity} remain identical, their entire subtrees are pruned. Only the modified assets (\texttt{Process()} and \texttt{Review()}) enter the drill-down queue.

In \textbf{Phase 2: Vertical Drill-down (Lines 5-20)}, the engine iterates over \(V_{mod}\) (Line 5) and invokes the \texttt{DrillDown} function. First, it loads and compares the scalar properties (Lines 8-9). If a property change is detected, it is recorded as a \textbf{root cause} (Line 11), such as updating the rule condition from \texttt{score >= 60} to \texttt{score >= 70} in Figure \ref{fig:approach}(b). 
Next, the function iterates over the union of old and new children (Line 14). If a child exists only in the new version, it is recorded as \textbf{ADDED} (Line 15-16); if it exists only in the old version, it is recorded as \textbf{DELETED} (Line 17-18). For children present in both versions but with different hashes (Line 19), the engine recursively invokes \texttt{DrillDown} (Line 20), continuing the downward traversal (e.g., from \texttt{Message} to \texttt{DecisionTable}). This recursion terminates when all direct property changes and structural additions/deletions are mapped, yielding the \textbf{causal diff tree} illustrated in Figure \ref{fig:approach}(c).

\textbf{3.3 Deterministic Context Injection.} We deterministically construct the agent's task specification based on the computed causal boundary. \textbf{Figure \ref{fig:approach}(c)} illustrates this injection for two affected tasks (\texttt{Process()} and \texttt{Review()}). For each task, the injector synthesizes a structured blueprint containing: (1) \textbf{Incremental Development Constraints}: a unified set of guidelines ensuring the agent only modifies the relevant scoped blocks; (2) \textbf{Causal Rationale}: the rendered causal diff tree showing exactly which rules changed and why; and (3) \textbf{Local Context}: the existing code file (e.g., \texttt{process.cpp}) alongside the newly fetched design assets. 

\textbf{3.4 Constrained LLM Code Evolution.} The LLM agent consumes the injected blueprint to evolve the code. \textbf{Figure \ref{fig:approach}(d)} demonstrates the execution: guided by the causal rationale (e.g., ``R1 changed, R2 added''), the agent modifies the exact logical blocks in \texttt{process.cpp}. Notably, because the \texttt{DecisionTable} (e.g., \texttt{EligibilityRules}) is a shared resource referenced by multiple sequence diagrams, multiple code files simultaneously receive the incremental modification requirements. The rigid boundary prevents the LLM from hallucinating unintended modifications, ensuring global consistency across all dependent files.

\textbf{3.5 Interactive KG Retrieval as a By-Product.} Beyond automated evolution, the KG serves as a queryable knowledge base for designers (Figure \ref{fig:approach}(e)). We equip a schema-aware agent with an interactive toolset—including schema exploration (\texttt{explore\_schema}), intent analysis, parameterized template queries, and multi-step Cypher generation (\texttt{query\_kg}). This allows designers to ask natural language questions (e.g., \textit{``Which flows use EligibilityRules?''}) and retrieve exact evidence paths, unlocking the static design assets as a dynamic, interactive engineering asset.

Finally, we report the feasibility of our Diff engine, code evolution, and KG Q\&A in the Early Results panel below.

\begin{figure}[htbp] 
\centering
\begin{tcolorbox}[colback=black!3!white,colframe=black!40!white,title=\textbf{Early Results from Industrial Deployment}, width=\columnwidth]
\textbf{Diff Engine:} We executed 3 mutation operations (add, modify, delete) across all KG node types.
The engine successfully captured 100\% of the multi-hop causal chains and generated bounded contexts. \\
\textbf{Code Evolution:} Given the injected diff, the LLM agent completed multi-file modifications on \textbf{10+} real-world sequence diagrams, achieving expected changes with zero unintended modifications. \\
\textbf{KG Q\&A:} Deployed end-to-end. Designers queried multi-hop structural paths in natural language and received accurate evidence paths. \\
\textit{(Large-scale quantitative evaluation remains future work.)}
\end{tcolorbox}
\vspace{-2em}
\end{figure}

\section{Future Plans}

While our early results demonstrate feasibility, the evaluation remains qualitative. Our future work will focus on quantitative evaluation and broader software evolution applications.

\textbf{A. Comprehensive Benchmark.} The absence of a standard public dataset for design-to-code evolution is a critical gap. We plan to construct a comprehensive benchmark by designing a suite of \textbf{multi-level mutation operators}. We will perform stratified sampling across all KG node types, injecting both \textbf{Atomic Operations} (add/modify/delete on isolated nodes) and \textbf{Combined Mutations} (multi-node linkage changes) to simulate complex scenarios. By tracing cascaded mutations, we can derive the precise \textit{Ground Truth} for evaluation.

\textbf{B. Quantitative Evaluation.}
Using the above benchmark, we will quantitatively evaluate our system against baselines including \textbf{LLM autonomous retrieval (e.g., agentic RAG)} and full-context prompting. 
We will measure the Diff engine's \textbf{Boundary Precision and Recall} (accuracy in capturing affected files) and \textbf{Evolution Safety} (compilation pass rate and unintended breaking changes).

\textbf{C. Generalizing to Other Artifacts.} While this paper focuses on design assets, our content-addressed diffing is domain-agnostic. We plan to generalize it to other structured artifacts, such as \textbf{Database Schema Evolution} (cascading to SQL views) and \textbf{Infrastructure-as-Code} (computing the blast radius in microservice topologies). This will further assist LLM agents in dependency-heavy evolution tasks.

\section{Related Work}

Agentic software evolution~\cite{hoda2026toward,sun2026agentic} is constrained by how agents acquire context during incremental modification. Existing frameworks use iterative planning \cite{bairi2024codeplan} or graph retrieval \cite{li2025graphcodeagent,fedorov2025gracg}, discovering dependencies probabilistically. 
Aegis uses a DAG to route target code files to relevant architecture documents \cite{aegis2026}. Our work targets design-to-code evolution: our Merkle DAG propagates design changes to compute the exact code files that must be modified, injecting this boundary as a hard constraint.
LLM-assisted requirement diffing \cite{yue2026llm}, traceability-guided generation \cite{chen2026incrertl}, and knowledge-graph-augmented evolution \cite{kang2026kcoevo} attempt to handle structural changes, yet none computes a deterministic impact boundary. Data dependency-aware generation from enhanced UML sequence diagrams \cite{mao2025DDI} targets initial code synthesis but not subsequent design evolution. Agent benchmarks and frameworks \cite{jimenez2024swe,zhang2024autocoderover,zhang2023repocoder} further reveal the persistent difficulty of multi-file consistency without a structural contract.

\section{Conclusion}

We presented a deterministic structural constraint approach to resolve boundary ambiguity in LLM-driven software evolution. By computing causal impact boundaries via a Merkle DAG and injecting them into agents, we ensure safe, scoped code modifications. Preliminary industrial deployment confirms its feasibility, with future work focused on quantitative evaluation and generalization.

\clearpage
\bibliographystyle{IEEEtran}
\bibliography{ieee}{}

\end{document}